\documentstyle[aps]{revtex}
\newfont{\boldit}{cmbxti10 at 14 pt}
\begin{document}
\title{Structure of Sn isotopes beyond \mbox{\boldmath
$N=82$}}

\author{L. Coraggio, A. Covello, A. Gargano, and N. Itaco}

\address{Dipartimento di Scienze Fisiche, Universit\`a di Napoli
Federico II, \\
and Istituto Nazionale di Fisica Nucleare, \\
 Complesso
Universitario di Monte S. Angelo, Via Cintia, I-80126 Napoli, Italy}

\date{\today}

\maketitle

\begin{abstract}
We have performed shell-model calculations for  $^{134,135}$Sn using a realistic effective interaction derived from the CD-Bonn nucleon-nucleon potential.  Comparison shows that the calculated results for $^{134}$Sn are in very good agreement with recent experimental data. This supports confidence in our predictions of the hitherto unknown spectrum of $^{135}$Sn.
\end{abstract}
\draft


The Sn isotopes play an essential role in the shell-model description of nuclear structure, as they provide the opportunity to study the change of nuclear properties when varying the number of neutrons over a large range. From the experimental point of view, there is information for practically all isotopes with valence neutrons in the 50-82 shell, namely from doubly magic $^{132}$Sn down to doubly magic $^{100}$Sn. It is therefore of great relevance to try to go beyond $^{132}$Sn, so as to test the shell-model predictions when having valence neutrons in the 82-126 shell. While this is not an easy task, since very neutron-rich Sn isotopes lie well away from the valley of stability, two very recent studies \cite{zhang97,korgul00} have provided some experimental information on $^{134}$Sn with two neutrons outside the doubly magic $^{132}$Sn core. 

During the past few years, we have studied several nuclei around $^{132}$Sn \cite{cov01} in terms of the shell model employing realistic effective interactions derived from modern nucleon-nucleon ($NN$) potentials.
We have therefore found it interesting to perform calculations of this kind for $^{134}$Sn and also to predict the spectroscopic properties of $^{135}$Sn with three neutrons in the 82-126 shell, for which no experimental spectroscopic data are available as yet.
In this paper, we present the results of our realistic shell model calculations starting with a brief description of how they have been obtained.

We consider $^{132}$Sn as a closed core and let the valence neutrons occupy the six single-particle levels $h_{9/2}$, $f_{7/2}$, $f_{5/2}$, $p_{3/2}$, $p_{1/2}$, and $i_{13/2}$ of the 82-126 shell.  The effective two-body matrix elements for the chosen model space have been derived from the CD-Bonn free $NN$ potential \cite{mach01}. 
This derivation has been performed within the framework of a $G$-matrix folded-diagram formalism, including
renormalizations from both core polarization and folded diagrams.
A detailed description of our derivation is given in  Ref.  \cite{cov97}. We should point out, however, that the effective interaction
used in the present work has been obtained including $G$-matrix diagrams through third order. 

Let us now come to the single-particle (SP) energies. The value of $\epsilon_{f_{7/2}}$ has been fixed at 
-2.455 MeV, as determined from the experimental one-neutron separation energy for $^{133}$Sn \cite{fogelberg99}. As
regards the other five SP energies , our adopted values (relative to the $f_{7/2}$ level) are (in MeV):
$\epsilon_{p_{3/2}}= 0.854$, 
$\epsilon_{h_{9/2}}= 1.561$, 
$\epsilon_{p_{1/2}}= 1.656$,
$\epsilon_{f_{5/2}}= 2.055$, and
$\epsilon_{i_{13/2}}= 2.694$.
The positions of the three single-neutron states  $p_{3/2}$, $h_{9/2}$, and $f_{5/2}$ are those determined 
in Ref. \cite{hoff96} by measuring $\gamma$ rays in coincidence with delayed neutrons following the decay of
$^{134}$In. In that work, it was also tentatively proposed the $p_{1/2}$ assignment to 
the state at 1.656 MeV. 
The energy of the $i_{13/2}$ level has been taken from Ref. \cite{urban99},
where it was estimated from the position of the 2.434 MeV level in $^{134}$Sb assumed to be
a $10^+$ state of 
$\pi g_{9/2}\nu i_{13/2}$ nature. 
Following the same procedure, we have calculated  
$\epsilon_{i_{13/2}}$ using an effective neutron-proton interaction derived from
the CD-Bonn potential. Our result comes very close to that of
\cite{urban99}.

We now present the results of our calculations. The spectrum of $^{134}$Sn
proposed in the experimental studies of Refs. \cite{zhang97,korgul00} is compared with
the  theoretical one in Fig. 1. In Tables I and II we report the 
calculated excitation energies up to about 2.8 and 4.0 MeV for the positive- 
and negative-parity states, respectively. We hope that these predictions may provide
a guidance to the interpretation of the results of  future 
studies on $^{134}$Sn. 

From the structure of our wave functions, it
turns out that the 20 states of Table I essentially arise from the 
configurations $(f_{7/2})^2$, $f_{7/2}p_{3/2}$, $(p_{3/2})^2$, 
$f_{7/2}h_{9/2}$, and $f_{7/2}p_{1/2}$.
The four members of the $(f_{7/2})^2$ multiplet all lie at an excitation 
energy smaller than 1.05 MeV and are separated  from the other states by a 
pronounced gap (about 500 keV), above which we predict the existence of 
four seniority-two
and one seniority-zero states coming from the $f_{7/2}p_{3/2}$ and $(p_{3/2})^2$
configurations, respectively. Just above these states we find the lowest- and
highest-spin members ($1^+$ and $8^+$) of the $f_{7/2}h_{9/2}$ multiplet.
All the states up to 2.4 MeV excitation energy are dominated by a
unique configuration, the percentage of configurations other than the dominant 
one reaching at most 23\% in the $2^+$ state at 1.501 MeV. The other members of 
$f_{7/2}h_{9/2}$ and $(p_{3/2})^2$ multiplets as well as the two states
arising from the $f_{7/2}p_{1/2}$ configuration are predicted in the energy region
above 2.4
MeV. It should be mentioned that for some of these states a significant admixture 
of these three configurations is present in our calculated wave functions.
As regards the negative-parity states, we have reported in Table II all the 8 
states arising from the $f_{7/2}i_{13/2}$ configuration. They are practically 
pure, the
percentage of this configuration ranging from 87 to 100\%.

Coming back to the comparison with experiment,
we see from Fig. 1 that the calculated  level scheme is in good agreement  with
the proposed one, supporting the
interpretation given in Refs. \cite{zhang97,korgul00}.    
Our calculated value of the ground-state binding energy relative to $^{132}$Sn is  
$5.986 \pm 0.064$ MeV, to be compared with the experimental one
$6.365 \pm 0.104$ MeV \cite{fogelberg99}. Note that the error on the calculated value arises from
the experimental error on the neutron separation energy of $^{133}$Sn.  
 
As far as the electromagnetic observables are concerned, only the $B(E2;6^{+} \rightarrow 4^{+})$ is
known with a measured value of $0.88 \pm 0.17$ W.u. \cite{zhang97}. To reproduce this experimental
$E2$ transition rate an effective neutron charge of $0.70 \pm 0.06 e$ is needed in our calculation.
This value is significantly smaller than $1.01e$, which was determined by Zhang {\em et al.} in \cite{zhang97}, 
but comes close to
the value $0.62e$ of Ref. \cite{sarkar01}. This 
is essentially due to the fact that 
in \cite{zhang97} the $4^+$ and $6^+$ states are interpreted as pure $(f_{7/2})^2$ states. 
By making use of $e^{\rm eff}_{n}=0.70e$,
we have calculated the $E2$ rates for transitions involving all the observed states. The
values are reported in Table III.   

Shell-model calculations  on $^{134}$Sn have been performed in some
previous studies \cite{korgul00,sarkar01,chou92}, where use was made of a
two-body interaction constructed by Chou and Warburton
\cite{chou92} starting from the Kuo-Herling effective interaction for the $^{208}$Pb
region \cite{herl72}.
A comparison between experimental and calculated spectra is shown in Refs. \cite{korgul00,sarkar01},
but not in \cite{chou92} since no experimental data where available at that time.
It should be mentioned, however, that very different conclusions are given in
\cite{korgul00} and \cite{sarkar01} as regards the adopted effective
interaction. In fact, while the comparison  between theory and experiment 
of Ref. \cite{korgul00}
shows a very good agreement, the calculated excitation energies reported in \cite{sarkar01} 
are all too high. Note that the results of Ref. \cite{sarkar01}
coincide with those of \cite{chou92}, both studies making use of the same set of SP energies,
which is different from that of Ref. \cite{korgul00}. However, we have verified that
the differences in the SP energies do not completely account for the results obtained in 
\cite{korgul00}. This point has already been evidenced in both \cite{korgul00} and \cite{sarkar01},
but no explanation was given.

Based on the good agreement between theory and experiment obtained for $^{134}$Sn, we have found
it interesting to predict some spectroscopic properties of $^{135}$Sn. This nucleus is expected to 
have a $\frac{7}{2}^-$ ground state \cite{korgul01,shergur01} with a binding energy of $8.393 \pm 0.401$ MeV
\cite{audi95}, as derived from systematic trend. No other experimental information
on its spectrum is presently available. In Table IV we report the calculated 
excitation energies 
of $^{135}$Sn up to about 1.5 MeV. Our 
calculations confirm the $\frac{7}{2}^{-}$ nature of  the ground state and 
predict for the binding energy  
the value $8.396 \pm 0.078$ MeV. As regards the excited states, we find that
the six members of the $f_{7/2}^{3}$
multiplet all lie at an excitation energy smaller than 1.0 MeV. From Table IV it 
can be seen that in this energy region we predict the existence of a second $\frac{3}{2}^-$ state (0.643 MeV) of
seniority-one nature, which is dominated by the $f_{7/2}^{2}p_{3/2}$ configuration. 
All the levels between 1.0 and 1.5 MeV arise from the  $f_{7/2}^{2}p_{3/2}$ configuration with only
two exceptions, the $\frac{1}{2}^{-}$  and $\frac{9}{2}^{-}$ states at 1.221 and 1.331 MeV, respectively.
The latter is, in fact, the seniority-one state of the $f_{7/2}^{2}h_{9/2}$ configuration while
the former contains almost the same percentage of the $f_{7/2}^{2}p_{3/2}$ and $f_{7/2}^{2}p_{1/2}$
configurations.

In Fig. 2 we show the behavior of the experimental energies  
\cite{urban00,nndc} of  the 
lowest-lying states with $J^{\pi}= \frac{3}{2}^-$, $\frac{5}{2}^-$, 
$\frac{7}{2}^-$, $\frac{9}{2}^-$, $\frac{11}{2}^-$, and
$\frac{15}{2}^-$
in the $N=85$ isotones with $52 \leq Z \leq 64$,  
as well as our predictions for $Z=50$. 
From this figure we see  that our calculated energies 
are quite consistent with the systematic trends.
Note that they refer to states 
arising from the  $f_{7/2}^{3}$ configuration, 
which, as mentioned above,  we have found to be the lowest-lying ones.
In the heavier isotones the lowest-lying states with 
$J^{\pi}= \frac{3}{2}^-$, $\frac{5}{2}^-$, 
$\frac{7}{2}^-$, $\frac{11}{2}^-$, and
$\frac{15}{2}^-$ have been interpreted as having 
this  nature, whereas the $\frac{9}{2}^-$ states were supposed to
arise from the $f_{7/2}^{2}h_{9/2}$ configuration \cite{nndc}.  
However,  only for the $Z \geq 60$ isotones,
which were studied by means of transfer reaction experiments,  there 
is evidence of the nature of these states, while 
no information is available for the lighter isotones.
In this situation, one cannot exclude the possibility that 
the lowest-lying $\frac{9}{2}^-$ state changes its nature when 
approaching
the proton shell closure. In fact, as a result of our calculations we find 
that the $\frac{9}{2}^-$ state belonging to the $f_{7/2}^{2}h_{9/2}$ 
configuration and that arising 
from the $f_{7/2}^{2}p_{3/2}$ configuration
lie at about 600  and 400 keV above the 
first one, respectively. In this connection it is worth
mentioning that for $^{137}$Te, with two protons in the 
50-82 shell, we predict three 
$\frac{9}{2}^-$ states in a even smaller energy range (from about
0.6 to 1.0 MeV), the lowest-lying one containing almost the
same percentage of the three-neutron configurations 
$f_{7/2}^{3}$ and $f_{7/2}^{2}p_{3/2}$. As regards the second and third
$\frac{9}{2}^-$ states, they are dominated by the $f_{7/2}^{3}$ and 
$f_{7/2}^{2}h_{9/2}$ configurations, respectively. 
Two excited states with $J^{\pi}=\frac{11}{2}^-$ and
$\frac{15}{2}^-$ have been observed in $^{137}$Te (see Fig. 2) at 
0.608 and 1.141 MeV, respectively. 
These energies are very well reproduced by our calculations (0.597 and 1.072 MeV).

In summary, we have shown that our realistic effective interaction derived from the CD-Bonn $NN$ potential leads to a very good description of the experimental data recently become available for $^{134}$Sn. We are therefore confident in the predictive power of our calculations and hope that this work may stimulate further experimental efforts to gain information on the structure of Sn isotopes beyond $N=82$.

\acknowledgements
\noindent
{This work was supported in part by the Italian Ministero dell'Universit\`a 
e della Ricerca Scientifica e Tecnologica (MURST).
NI thanks the European Social Fund for financial support.}

\begin{figure}
\caption{Experimental and calculated spectrum of $^{134}$Sn.}
\end{figure}

\begin{figure}
\caption{Energy systematics for the levels of the 
$(f_{7/2})^{3}$ multiplet in the $N=85$ isotones. All  
energies are relative to the $\frac{7}{2}^{-}$ state. 
Data for $Z \geq 52$ (solid symbols) are taken from
[15,16] while for $Z=50$  our calculated values are reported (open symbols).}
\end{figure}

\newpage

\mediumtext
\begin{table}
\setdec 0.00
\caption{Calculated excitation energies   
for the positive-parity states of $^{134}$Sn up to 2.8 MeV.}

\begin{tabular}{rc}
J & $E$(MeV) \\
\tableline
$0$  & 0.0 \\
$2$  & 0.640 \\
$4$  & 0.936  \\
$6$  & 1.050  \\
$2$  & 1.501 \\
$4$  & 1.765 \\
$3$  & 1.890 \\
$5$  & 1.944 \\
$0$  & 2.167  \\
$1$  & 2.304  \\
$8$  & 2.382 \\
$2$  & 2.490  \\
$6$  & 2.512 \\
$4$  & 2.515  \\
$2$  & 2.545 \\
$3$  & 2.635 \\
$4$  & 2.703 \\
$5$  & 2.733 \\
$3$  & 2.740 \\
$7$  & 2.744 \\
\end{tabular}
\end{table}

\mediumtext
\begin{table}
\setdec 0.00
\caption{Calculated excitation energies  
for the negative-parity states of $^{134}$Sn up to about 4.0 MeV.}

\begin{tabular}{rc}
$J$ & $E$(MeV) \\
\tableline
$3$  & 3.247 \\
$5$  & 3.621 \\  
$4$  & 3.684 \\
$7$  & 3.713 \\
$9$  & 3.752 \\
$6$  & 3.848 \\
$8$  & 3.925 \\
$10$ & 3.973   \\
\end{tabular}
\end{table}

\mediumtext
\begin{table}
\setdec 0.00
\caption{Calculated $B(E2)$ (in W.u.)  
in $^{134}$Sn.}
\begin{tabular}{cl}
$J^{\pi}_{i} \rightarrow J^{\pi}_{f}$  & $B(E2)$ \\
\tableline
$2^{+} \rightarrow 0^{+}$ & 1.72 \\
$4^{+} \rightarrow 2^{+}$ & 1.71 \\
$6^{+} \rightarrow 4^{+}$ & 0.88 \\
$8^{+} \rightarrow 6^{+}$ & 0.12 \\
\end{tabular}
\end{table}

\mediumtext
\begin{table}
\setdec 0.00
\caption{Calculated excitation energies   
for states of $^{135}$Sn up to about 1.5 MeV.}

\begin{tabular}{rc}
$J^{\pi}$ & $E$(MeV)\\
\tableline
$\frac{7}{2}^{-}$  & 0.0 \\
$\frac{5}{2}^{-}$  & 0.226 \\
$\frac{3}{2}^{-}$  & 0.356 \\
$\frac{11}{2}^{-}$  & 0.611 \\
$\frac{3}{2}^{-}$  & 0.643  \\
$\frac{9}{2}^{-}$  & 0.706 \\
$\frac{15}{2}^{-}$  &0.911 \\
$\frac{9}{2}^{-}$  & 1.093 \\      
$\frac{7}{2}^{-}$  & 1.192 \\     
$\frac{1}{2}^{-}$  & 1.221 \\      
$\frac{3}{2}^{-}$  & 1.264 \\      
$\frac{5}{2}^{-}$  & 1.295 \\      
$\frac{7}{2}^{-}$  & 1.298 \\      
$\frac{9}{2}^{-}$  & 1.331 \\      
$\frac{5}{2}^{-}$  & 1.430 \\    
\end{tabular}
\end{table}

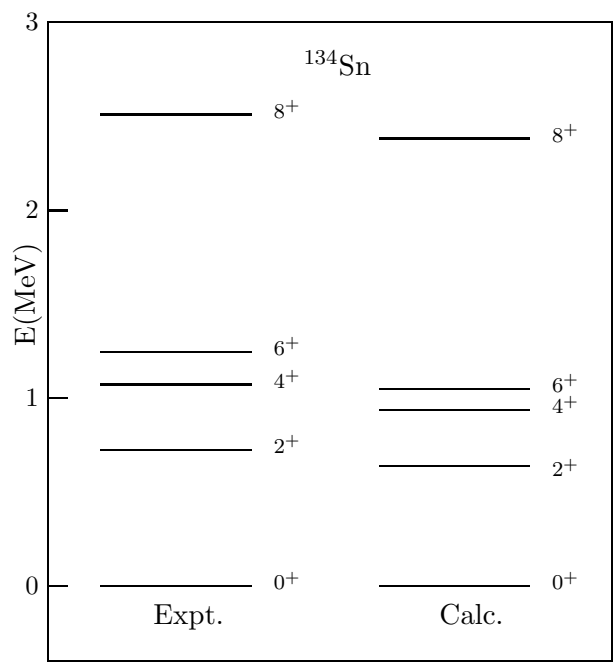
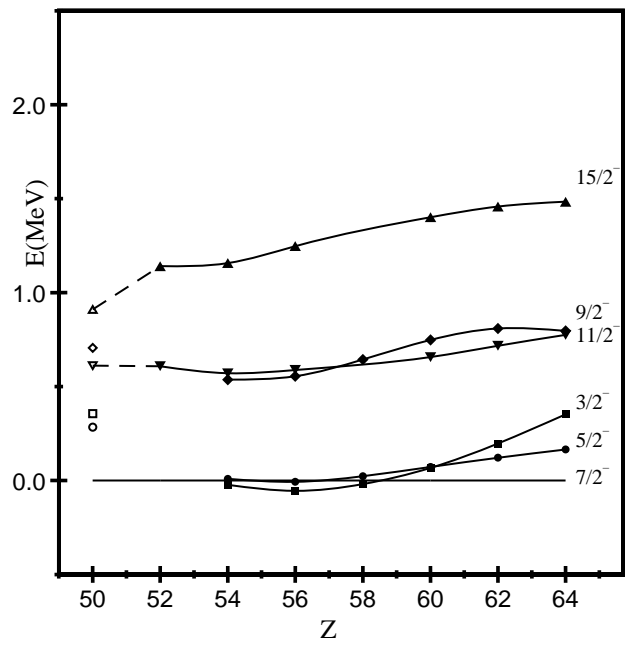
\end{document}